\documentclass[11pt]{article}
\usepackage{amsmath}
\usepackage{amsfonts}
\usepackage{amssymb}
\usepackage{graphicx}

\def\le{\leq}
\def\ge{\geq}

\newcommand{\ls}[1]
   {\dimen0=\fontdimen6\the\font \lineskip=#1\dimen0
\advance\lineskip.5\fontdimen5\the\font \advance\lineskip-\dimen0
\lineskiplimit=.9\lineskip \baselineskip=\lineskip
\advance\baselineskip\dimen0 \normallineskip\lineskip
\normallineskiplimit\lineskiplimit \normalbaselineskip\baselineskip
\ignorespaces }

\newtheorem{theorem}{Theorem}

\newcommand {\dfn} {\stackrel{\Delta} {=}}

\newcommand {\reals} {{\rm I\!R}}

\newcommand {\bk} {\mbox{\boldmath $k$}}

\newcommand {\bK} {\mbox{\boldmath $K$}}

\newcommand {\hk} {\hat{k}}

\newcommand{\calU}{{\cal U}}
\newcommand{\calV}{{\cal V}}

\newcommand{\calX}{{\cal X}}
\newcommand{\calY}{{\cal Y}}

\begin{document}
\thispagestyle{empty}
\title{On the Shannon Cipher System with a Capacity--Limited Key--Distribution Channel}
\author{Neri Merhav}
\date{}
\maketitle

\begin{center}
Department of Electrical Engineering \\
Technion - Israel Institute of Technology \\
Haifa 32000, ISRAEL \\
{\tt merhav@ee.technion.ac.il}
\end{center}
\vspace{1.5\baselineskip}
\setlength{\baselineskip}{1.5\baselineskip}

\begin{abstract}
We consider the Shannon cipher system in a setting where
the secret key is delivered to the legitimate receiver 
via a channel with limited
capacity. For this setting, we characterize 
the achievable region in the space of three figures
of merit: the security (measured in terms of the
equivocation), the compressibility of the cryptogram, and the distortion
associated with the reconstruction of the plaintext source. Although lossy reconstruction
of the plaintext does not rule out the option that the (noisy) decryption key would
differ, to a certain extent, from the encryption key, we show, nevertheless, that the
best strategy is to strive for perfect match between the two keys,
by applying reliable channel
coding to the key bits, and to control the distortion solely via rate--distortion coding of
the plaintext source before the encryption. In this sense, our result has a flavor similar to that of the
classical source--channel separation theorem. Some variations and extensions of this
model are discussed as well.
 
\vspace{1cm}

\end{abstract}

\section{Introduction}

In the classical Shannon--theoretic approach to cryptology (see, e.g., 
\cite{Shannon49},\cite{Massey88},\cite{Yamamoto91} and references therein),
two assumptions are traditionally made. The first is that the reconstruction of the
decrypted plaintext source at the legitimate receiver is distortion--free (or almost
distortion--free), and the second, which is related, is that the encryption and the
decryption units share identical copies of the same key. Yamamoto \cite{Yamamoto97} has
relaxed the first assumption and extended the theory of Shannon secrecy systems into
a rate--distortion scenario, allowing lossy reconstruction at the legtimate receiver.

In this correspondence, we examine also the second assumption. 
Referring to Fig.\ 1,
we consider the case where the key is delivered to the legitimate receiver
across a channel, which is cryptographically secure, but has 
limited capacity. 
For this setting, we characterize
the achievable region in the space of three figures
of merit: the security level (measured in terms of the
equivocation), the compressibility of the cryptogram, and the distortion
associated with the reconstruction of the plaintext source.

One conceptually simple approach to handle 
such a situation would be to apply a reliable
channel code to the encryption 
key bits, at a rate below the capacity of the channel, and
thereby obtain, with high probability, the exact copy of 
the transmitted key bits at the receiver
side. With this approach, however, the effective 
key rate, and hence the security level
in terms of the equivocation, is limited by the channel capacity. The question that naturally
arises at this point, especially in the lossy reconstruction scenario,
is whether this is the best one can do.

To sharpen the question, let us even assume that there is an unlimited
reservoir of random key bits at the transmitter side, denoted 
$\bK=(K_1,K_2,\ldots)$, $K_i\in\{0,1\}$, $i=1,2,\ldots$. Then, 
perhaps one might
wish to use more key rate (somewhat above capacity) 
for encryption and thereby increase the security of the cryptogram
at the expense of some distortion at the reconstruction, due to the unavoidable
mismatch between the encryption and decryption keys.
To explore this point, let us consider a few speculative strategies.

In the first strategy, one sends the 
key bits $\bK$ across the channel uncodedly (assuming, for
simplicity, that the channel has a binary input--output alphabet). 
Referring to Fig.\ 1, let us take then
$N=n$ and $X_i=K_i$, $i=1,2,\ldots$. In this case,
the noisy version of the key, obtained at the receiver side, 
$K_i'=Y_i$, is of course somewhat different
from the original key. However, since only lossy reconstruction of the
plaintext is required at the receiver side, 
it may seem conceivable that a reasonably small difference
between the keys at both ends could be managable 
and thus cause a reasonably small distortion 
in the reconstruction. This is relatively easy to have
if the encryption of the source precedes 
compression, as proposed in \cite{JIPSR04}:
One may apply, for example, a certain memoryless mapping 
from the key bit stream into 
a stream of symbols $Z_1,Z_2,\ldots$ 
taking (two of the) values in the alphabet of plaintext source,
$\calU$. Then assuming that $\calU$ is a 
commutative group endowed with an addition operation $\oplus$
(e.g., addition modulo the alphabet size), 
one can create the enctypted sequence $U_i'=U_i\oplus Z_i$, $i=1,2,\ldots$
and then compress 
the block $(U_1',\ldots,U_n')$ with
$(K_1',\ldots,K_n')$ as side information 
at the receiver, using a Slepian--Wolf
encoder \cite{SW73} in the lossless case, or a 
Wyner--Ziv code \cite{WZ76} in the lossy case. Assuming, for simplicity,
lossless compression, then upon decompressing 
the source at the receiver side and
obtaining $(\tilde{U}_1,\ldots,\tilde{U}_n)$ 
(which is with high probability equal to
$(U_1',\ldots,U_n')$), one `subtracts' the noisy version of the key and obtain
(with high probability) the 
reconstruction $V_i=U_i'\ominus Z_i'$, $i=1,2,\ldots$,
where $Z_i'$ is the corresponding noisy version of $Z_i$.
Now, since $V_i\ominus U_i=Z_i\ominus Z_i'$, for all $i$,
then for a difference distortion measure 
$d(U_i,V_i)=\rho(V_i\ominus U_i)$, the distortion
between $U_i$ and its reconstruction $V_i$ 
is identical to the distortion between
the original key $Z_i$ and its noisy version $Z_i'$.

A somewhat more sophisticated version of this scheme 
generates $Z_1,Z_2,\ldots$ from the key bits 
using a simulator of a certain (memoryless) process
(see, e.g., \cite{SV96} and references therein), and then applies
a good source--channel code to
encode $(Z_1,\ldots,Z_n)$ across the channel. The reconstructed version 
at the receiver side, $Z_1',Z_2',\ldots$,
would then have the minimum possible distortion
relative to $(Z_1,\ldots,Z_n)$,
given by the distortion--rate function of 
$\{Z_i\}$ computed at the channel capacity, and therefore so
would be also the distortion between $\{U_i\}$ 
and $\{V_i\}$. Moreover, there is an additional degree
of freedom with regard to the choice of 
the probability law of $\{Z_i\}$ for trading off between
the security, which is given by the entropy 
rate of $\{Z_i\}$, and the distortion, i.e.,
distortion--rate function of $\{Z_i\}$ computed at the channel capacity.

Another solution strategy 
may be based on the following point: 
Note that for the purpose of reliable transmission
and decoding of the key bits across the channel, 
the cryptogram (denoted by $W^m$ in Fig.\ 1),
which is a function of these key
bits as well, may serve as useful side 
information at the decoder, unless it is statistically
independent of these bits. Thus, one would
speculate that it might be wise to allow some dependence
between $W^m$ and $\bK$ and thus sacrifice 
some compression performance at the benefit 
gaining performance in communicating the key across the channel.
Let us assume that
the bits of the key string $K^m=(K_1,\ldots,K_m)$ 
are XORed (added modulo 2) with the bits of the compressed
version of the source. Then, if the compression algorithm
is designed in such a way such the bits of the compressed version of $U^N$
are {\it not} symmetric, then $W^m$ 
is correlated to 
$K^m$, and so $W^m$ can be viewed as a noisy version of $K^m$,
which was transmitted uncodedly
across a ``parallel channel''.
In such a case, we can then think of the key
bits as being encoded using a {\it systematic} 
code across the combined channel whose outputs are
$W^m$ and $Y^n$ and the 
effective rate of this code is smaller than that
over the original channel depicted in Fig.\ 1. Another way to look at this
is the following: The key string $K^m$ can be compressed by a Slepian--Wolf
encoder given $W^m$ (as side information 
at the decoder) before being channel coded, thus increasing the
effective capacity by a factor given by the 
reciprocal of the conditional entropy of the
key given the cryptogram.

We show in this correspondence that none of 
the ideas raised in the last four paragraphs,
nor any other creative idea one may have, can work
better than the first strategy 
we mentioned earlier, which is the following: At
the lower part of the encoder of Fig.\ 1 
(the ``key encoder''), use a good channel code
at rate below capacity, whose role is to
reliably transmit a certain amount of key bits.
At the upper block of the encoder of Fig.\ 1,
first compress $U^N$ by an optimal rate--distortion
code to obtain $NR(D)$ bits, 
where $R(D)$ is the rate--distortion function of $U^N$,
and then encrypt the compressed bitstream with the same bits that are fed into the
channel code. At the receiver, first decode 
the key bits from the channel output,
and then use them to decrypt and decompress the source.

The result on the optimality of this scheme has a flavor similar to that of the
classical source--channel separation theorem in three
aspects: (i) There is a complete decoupling between
source coding (for $U^N$) and channel coding (for the key bits) 
from the operative point of view as well 
as from the viewpoint of code design (unlike in the other
strategies described above), (ii) 
the best possible strategy of controlling the
distortion is only via rate--distortion coding, 
and (iii) the necessary and sufficient 
condition for perfect secrecy
is $NR(D)\le nC$, which is of the same form as 
the source--channel separation theorem.

The outline of this correspondence is as follows. In Section 2, we define notation conventions and give
a formal definition of the problem. In Section 3, we state and prove the main result, and
in Section 4, we discuss a few variations and extensions.

\section{Notation Conventions and Problem Definition}

We begin by establishing some notation conventions.
Throughout this paper, scalar random 
variables (RV's) will be denoted by capital
letters, their sample values will be denoted by
the respective lower case letters, and their alphabets will be denoted
by the respective calligraphic letters.
A similar convention will apply to
random vectors and their sample values,
which will be denoted with same symbols superscripted by the dimension.
Thus, for example, $U^N$ ($N$ -- positive integer)
will denote a random $N$-vector $(U_1,...,U_N)$,
and $u^N=(u_1,...,u_N)$ is a specific vector value in $\calU^N$,
the $N$-th Cartesian power of $\calU$. 

Sources and channels will be denoted generically by the letter $P$, 
subscripted by the name of the RV and its conditioning,
if applicable, e.g., $P_U(u)$ is the probability function of
$U$ at the point $U=u$, $P_{Y|X}(y|x)$
is the conditional probability of $Y=y$ given $X=x$, and so on.
Whenever clear from the context, these subscripts will be omitted.
Information theoretic quantities like entropies and mutual
informations will be denoted following the usual conventions
of the Information Theory literature, e.g., $H(U^N)$, $I(X^n;Y^n)$,
and so on. For single--letter
information quantities (i.e., when $n=1$ or $N=1$), 
subscripts will be omitted, e.g., $H(U^1)=H(U_1)$ will
be denoted by $H(U)$, 
similarly, $I(X^1;Y^1)=I(X_1;Y_1)$ will be denoted by $I(X;Y)$, and so on.

We now turn to the formal description 
of the model and the problem setting,
as described in the Introduction, and referring to Fig.\ 1.
A source $P_U$, generates
a sequence of
independent copies, $U_1,U_2,\ldots$ of a finite--alphabet RV,
$U\in\calU$, whose entropy is $H(U)\dfn -\sum_{u\in\calU}P_U(u)\log_2P_U(u)$.
At the same time and independently,
a discrete memoryless channel (DMC) $P_{Y|X}$ receives input symbols
$x_1,x_2,\ldots$ 
with coordinates taking values in a finite alphabet $\calX$, and
produces output symbols $y_1,y_2,\ldots$
with coordinates taking values in a finite alphabet $\calY$, according to
a conditional probability law given by the product of single--letter
transition probabilities $\prod_t P_{Y|X}(y_t|x_t)$.
The relative rate between the operation of the channel $P_{Y|X}$ and that of the
source is $\lambda$ channel symbols per source symbol. This means that
while the 
source generates a block of $N$ symbols, say, $U^N=(U_1,\ldots,U_N)$,
according to the above mentioned probability law,
the channel conveys $n=\lambda N$ transmissions,\footnote{
Without essential loss of generality, 
we will assume that $\lambda N$ is a positive integer.}
i.e., it receives a channel
input block of length $n$, $X^n=(X_1,\ldots,X_n)$, and outputs another
block of the same length $Y^n=(Y_1,\ldots,Y^n)$ according to the
above described conditional probability law.
Let $C=\max_{P_X}I(X;Y)$ denote the channel capacity.

In addition to the source $P_U$ and 
the channel $P_{Y|X}$, yet another source, $P_K$,
henceforth referred to as the {\it key source},
generates an infinite sequence of i.i.d.\ purely random bits,
$\bK=(K_1,K_2,\ldots)$, independently of the source $U_1,U_2,\ldots$.
The operation rate of the key source
relative to the source $P_U$ (and the channel $P_{Y|X}$) will be immaterial,
i.e., we will assume that the reservoir of key bits, for every finite
period of time, is sufficiently large so that it is effectively unlimited.

A {\it block code for joint coding and
encryption} with parameters $n$ and $\lambda=n/N$,
consists of three mappings. The first mapping is the 
compressor--encrypter
$f_N:\calU^N\times\{0,1\}^\infty\to\{0,1\}^m$, where $m=\mu N$,
$\mu > 0$ being the compression rate. 
Upon receiving a source vector $u^N\in\calU^N$ and a key sequence
$\bk\in\{0,1\}^\infty$,
this mapping produces a binary cryptogram
$w^m\in\{0,1\}^m$
according to $w^m=f_N(u^N,\bk)$. The second mapping
is the key--encoder $g_N:\{0,1\}^\infty\to {\calX}^n$, which produces a channel
input vector $x^n$ according to $x^n=g_n(\bk)$. 
Finally, the third mapping is the decoder 
$h_N:\{0,1\}^m\times{\calY}^n\to{\calV}^N$, where $\calV$ is the
reproduction alphabet. Upon receiving a cryptogram $w^m$ and a channel
output vector $y^n$, the decoder produces a reproduction vector
according to $v^N=h_N(w^m,y^n)$.

Let $d:\calU\times\calV\to \reals^+$ denote
a single--letter distortion measure between 
source symbols and the reproduction symbols,
and let the distortion between the vectors, $u^N\in\calU^N$ and $v^N\in\calV^N$,
be defined additively across the corresponding components, as usual. 
We will assume that $d$ is bounded, i.e., 
$d_{\max}\dfn \max_{u,v}d(u,v) < \infty$.
Let $R(D)$ denote the rate--distortion function of the source $P_U$
with respect to $d$.

An $(n,\lambda,D,R_c,h)$ code 
for joint coding and encryption
is a block code 
with parameters $n$ and $\lambda$, 
as above, which also satisfies the following requirements:
\begin{itemize}
\item [1.] The expected distortion 
between the source and the reproduction satisfies
\begin{equation}
\sum_{i=1}^N Ed(U_i,V_i)\le ND.
\end{equation}
\item [2.] The rate of the cryptogram
satisfies
\begin{equation}
\mu= \frac{m}{N}\le R_c.
\end{equation}
\item [3.] The equivocation of the source satisfies
\begin{equation}
H(U^N|W^m)\ge Nh.
\end{equation}
\end{itemize}

For a given $\lambda$, a triple $(D,R_c,h)$ is said to be {\it achievable}
if for every $\epsilon > 0$, there is a sufficiently large $n$
for which 
$(n,\lambda,D+\epsilon,R_c+\epsilon,h-\epsilon)$ 
block codes for joint coding and encryption exist.
Our purpose, in this paper is to characterize 
the {\it achievable region} of triples 
$(D,R_c,h)$, i.e., the set of all achievable
triples $(D,R_c,h)$. 

\section{Main Result}

Our main coding theorem is the following:
\begin{theorem}
A triple $(D,R_c,h)$ is achievable if and only if the following conditions
are both satisfied:
\begin{itemize}
\item [(a)] $h \le h^*(D)\dfn 
H(U)-[R(D)-\lambda C]_+$, where $[a]_+\dfn\max\{a,0\}$.
\item [(b)] $R_c \ge R(D)$.
\end{itemize}
\end{theorem}

It should be noted that for a given $D$, there is no 
conflict (or interaction) between
maximizing $h$ and minimizing $R_c$: As is well known, $R_c$ is lower
bounded by $R(D)$ even if there is no security requirement, but on the
other hand, even in the presence of the highest 
possible security level requirement, of $h^*(D)$, the 
compression ratio $R(D)$ is still
achievable \cite{Yamamoto97}. By the same token, 
and as will be evident from the proof, $h$ is upper bounded by
$h^*(D)$ even if there is no compressibility requirement,
yet it remains achievable even 
if the compression ratio of $R(D)$ is required.

\noindent
The remaining part of this section is devoted to the proof of
Theorem 1.

\noindent
{\it Proof.} We begin with the converse part. Let
an $(n,\lambda,D+\epsilon,R_c+\epsilon,h-\epsilon)$ block code 
for joint coding and encryption be given. Now, since
\begin{equation}
h^*(D)=H(U)-[R(D)-\lambda C]_+=\min\{H(U),H(U)-R(D)+\lambda C\},
\end{equation}
we have to prove that both $h\le H(U)$ and $h\le H(U)-R(D)+\lambda C$.
The first bound is trivial since
\begin{equation}
\label{1st}
N(h-\epsilon)\le H(U^N|W^m)\le H(U^N)=NH(U),
\end{equation}
where the first inequality is by definition of
an $(n,\lambda,D+\epsilon,R_c+\epsilon,h-\epsilon)$ block code 
for joint coding and encryption. The inequality $h\le H(U)$ now
follows from the arbitrariness of $\epsilon > 0$.
As for the second bound, we have
\begin{eqnarray}
\label{2nd}
N(h-\epsilon) &\le& H(U^N|W^m)\nonumber\\
&=&H(U^N|W^m,Y^n)+I(U^N;Y^n|W^m)\nonumber\\
&=&H(U^N|W^m,Y^n,V^N)+H(Y^n|W^m)-H(Y^n|W^m,U^N)\nonumber\\
&\le&H(U^N|V^N)+H(Y^n)-H(Y^n|W^m,U^N,X^n)\nonumber\\
&=&H(U^N)-I(U^N;V^N)+H(Y^n)-H(Y^n|X^n)\nonumber\\
&\le&NH(U)-NR(D+\epsilon)+I(X^n;Y^n)\nonumber\\
&\le&N[H(U)-R(D+\epsilon)]+nC,
\end{eqnarray}
where the second line is 
a standard identity, the third is because $V^N$ is a function of
$(W^m,Y^n)$, the fourth is because conditioning reduces entropy (used thrice),
the fifth is due to the fact that $(U^N,W^m)\to X^n\to Y^n$ is a
Markov chain, the sixth is due to the memorylessness of the source
and the fact that $R(D)=\min\{I(U;V):~Ed(U,V)\le D\}$ (which is also convex),
and the last line is due to the memorylessness of the channel and
the fact that $C=\max_{P_X}I(X;Y)$. Again, 
dividing by $N$, and using the arbitrariness of $\epsilon > 0$
as well as the continuity of $R(D)$,
we get the second bound on $h$, and so,
the necessity of condition (a) follows.

The proof of the necessity of condition (b) is similar to the proof of
the converse to the 
ordinary rate--distortion coding theorem, except that the presence of
$Y^n$ (which is independent of $U^N$) 
at the decoder has to be taken into account: 
\begin{eqnarray}
N(R_c+\epsilon)&\ge&H(W^m)\nonumber\\
&\ge&H(W^m|Y^n)\nonumber\\
&\ge&I(U^N;W^m|Y^n)\nonumber\\
&=&\sum_{i=1}^N[H(U_i|U^{i-1},Y^n)-H(U_i|U^{i-1},W^m,Y^n)]\nonumber\\
&\ge&\sum_{i=1}^N[H(U_i)-H(U_i|W^m,Y^n)]\nonumber\\
&=&\sum_{i=1}^NI(U_i;W^m,Y^n)\nonumber\\
&\ge&\sum_{i=1}^NI(U_i;V_i)\nonumber\\
&\ge&NR(D+\epsilon),
\end{eqnarray}
where the first line is by definition of
an $(n,\lambda,D+\epsilon,R_c+\epsilon,h-\epsilon)$ block code 
for joint coding and encryption, the second, third, fourth and sixth
are standard identities and inequalities, the fifth is based
on the memorylessness of the source and its independence 
of $Y^n$, the seventh is based on the
data processing inequality and the fact that $V_i$ is a function
of $(W^m,Y^N)$, and the last inequality
is again by the informational definition of $R(D)$ and its convexity.
Taking again $\epsilon$ to zero, this completes the proof of the converse part of Theorem 1.

As for the direct part, consider the following (conceptually) simple coding
scheme. For a given arbirarily small $\epsilon > 0$, let 
$\ell=\min\{n(C-\epsilon),N[R(D)+\epsilon]\}$
and let $x^n=g_N(k_1,\ldots,k_\ell)$ be given by a channel code whose
error probability is below some $\delta > 0$, provided that $n$
is sufficiently large. Since the rate of this code never exceeds $C-\epsilon$,
such a channel code exists by the classical channel coding theorem.
As for $f_N$, first apply a rate--distortion
code for $U^N$, whose rate is $R_c=R(D)+\epsilon$, and then encrypt
$\ell$ of the resulting $m=N[R(D)+\epsilon]$ bits by $(k_1,\ldots,k_\ell)$
(using the ordinary bit--by--bit XOR). As for the equivocation, we have
\begin{eqnarray}
H(U^N|W^m)&=&H(U^N)-I(U^N;W^m)\nonumber\\
&=&NH(U)-H(W^m)+H(W^m|U^N)\nonumber\\
&\ge&NH(U)-N[R(D)+\epsilon]+H(W^m|U^N)\nonumber\\
&=&NH(U)-N[R(D)+\epsilon]+\ell\nonumber\\
&=&NH(U)-N[R(D)+\epsilon]+\min\{n(C-\epsilon),N[R(D)+\epsilon]\}\nonumber\\
&\ge&N\left(H(U)-[R(D)-\lambda C]_+-2\epsilon\max\{1,\lambda\}\right),
\end{eqnarray}
where the first inequality follows from the fact that
the rate--distortion code is at rate $R(D)+\epsilon$, and the
following equality is due to the fact that $\ell$ bits of the
compressed bit string are encrypted. At the decoder, first, the
$\ell$ key bits $(k_1,\ldots,k_\ell)$ are decoded, and then
the decoded key bits $(\hk_1,\ldots,\hk_\ell)$ are used to
decrypt $w^m$ and then use the rate--distortion decoder to produce $v^N$.
With probability at least $1-\delta$, the decoded key bits
$(\hk_1,\ldots,\hk_\ell)$ agree with the original ones $(k_1,\ldots,k_\ell)$
and then $w^m$ is decrypted correctly to produce the appropriate
reproduction vector $v^N$ within distortion $D$. At the event of erroneous
decoding of $(k_1,\ldots,k_\ell)$, the distortion can only be bounded by
$d_{\max}$, but this should be weighed by the probability of error, which
is upper bounded by $\delta$, and hence contributes only an arbitrarily small
additional distortion. This completes the proof of Theorem 1.

\section{Discussion}

In this section, we discuss a few variations and extensions of the
model considered. 

\subsection{Source--Channel Separation}

We have already mentioned in the Introduction that Theorem 1 has
the spirit of a separation theorem, from several points of view.
Among them is the immediate observation that perfect security
(in the sense that $h=H(U)$) can be achieved if and only if $R(D)\le \lambda C$,
an inequality of the very same form as that of the classical joint
source--channel separation theorem. 
In this context, we should also point out that it is straightforward to extend
our setup to a situation of ordinary joint source--channel coding,
corresponding to the case where the cryptogram $w^m$ needs to be transmitted
via a noisy channel, independent of the key distribution channel. The only
modification to Theorem 1 would be to replace $R_c$ in part (b) by
the capacity of the main channel. Thus, we have a two--fold separation theorem.

\subsection{Simple Coding and Decoding in Special Cases}

Suppose that the compressibility
of the cryptogram is not an issue, in other words, $R_c$ is immaterial
and we are only interested in the tradeoff between $D$ and $h$. 
In this case, there exist situations where optimal performance can be
achieved using very simple coding systems, 
similiarly to the well--known special cases, where this can be done
in the context of classical joint source--channel coding 
(see, e.g., \cite{Gastpar03}).
Let us suppose, for example, that $\calU=\calX=\calY=\calV$,
$\lambda=1$, and that the distortion measure 
$d$ is a difference distortion measure,
i.e., $d(u,v)=\rho(v\ominus u)$ for a well--defined subtraction operation
(cf.\ the corresponding
discussion in the Introduction). Suppose further that $P_U$,
which is the uniform distribution over $\calU$,
is the capacity--achieving input for the channel $P_{Y|X}$ and that 
$P_{Y|X}$ in turn achieves the rate--distortion function of $P_U$ at 
distortion level $D$, i.e., $R(D)=C$. For example, $P_U$ may be the BSS and
$P_{Y|X}$ may be the BSC with crossover probability $D$.
Then one can easily achieve perfect secrecy,
$h=H(U)=\log|\calU|$, at the minimum possible 
distortion, i.e., $D=R^{-1}(C)$ ($R^{-1}(\cdot)$ being
the distortion--rate function of $U$) in the following manner,
which is similar to one of the strategies discussed in the Introduction:
Let $Z_1,Z_2,\ldots$ be a 
simulated memoryless process, generated from $\bK$, with
the same (uniform) distribution 
as $U_1,U_2,\ldots$. Note that when $|\calU|$ is
a power of 2, this is very easy to implement since $U$ is uniform.
For encryption, let
$W_i=U_i\oplus Z_i$. Then, obviously, $H(U^N|W^N)=NH(U)=N\log|\calU|$
since $U_i$ and $Z_i$ are uniformly distributed and independent, and so,
perfect secrecy is guaranteed. As for the key transmission, let us
send $\{Z_i\}$ uncodedly across the channel, i.e., $X_i=Z_i$.
Since $P_{Y|X}$ achieves
the rate--distortion of $\{U_i\}$, and hence also that of $\{Z_i\}$,
then the channel output $\{Y_i\}$ will have distortion $D$ relative to
$\{Z_i\}$. At the the decoder, we simply apply the equation
$V_i=W_i\ominus Y_i$. Since $V_i\ominus U_i=Z_i\ominus Y_i$, then
$Ed(U_i,V_i)=E\rho(V_i\ominus U_i)=E\rho(Z_i\ominus Y_i)=D=R^{-1}(C)$. Thus,
optimal performance is achieved using a very simple system once we
have an independent copy of $\{U_i\}$ as a key.

\subsection{A Wider Class of Joint Encoders}

Another point regarding the case
where $R_c$ is immaterial, is the following: It turns
out that part (a) of Theorem 1 (both the necessity and the sufficiency)
would still apply even if we 
broaden the scope to a wider
class of encoders that allow both $x^n$ and $w^m$ to depend on both
$u^N$ and $\bk$.
This means that $g_N$ is redefined as $g_N:\{0,1\}^\infty\times\calU^N\to
\calX^n$, and so, $x^n=g_N(\bk,u^N)$. The direct part would use the
same scheme as before. As for the converse part, note that eq.\ (\ref{2nd})
is general enough to allow this setup.
The conclusion then is that if only $D$ and $h$ are
the figure of merits of interest, then a good key code $g_N$ need not
really use its accessibility to $u^N$. 
The situation becomes somewhat more involved when the compressibility
is brought back into the picture, because 
then the encoder has two paths through which
it can pass descriptions of the source.
Note that if $R(D)\le \lambda C$, the encoder can transmit the entire
description via the key distribution channel, without using the main
channel at all, thus $R_c=0$.

\subsection{Securing the Reproduction Sequence}

Consider the case where one is interested not only to
guarantee a certain security level $h$ with regard to the original source,
but also to guarantee a security level $h'$ with regard to the
reproduction $V^N$. This makes sense because it is actually $V^N$
the part of the information that is communicated to the legitimate
receiver and thus has to be protected (see also \cite{Merhav05}). To derive necessary conditions
for securing $V^N$ at level $h'$, we consider two chains of inequalities.
The first is the following:
\begin{eqnarray}
N(h'-\epsilon)&\le& H(V^N|W^m)\nonumber\\
&\le& H(V^N)\nonumber\\
&\le& \sum_{i=1}^N H(V_i)\nonumber\\
&=& N H(V|J)\nonumber\\
&\le& N H(V)
\end{eqnarray}
where $J$ is random variable taking values in the set $\{1,\ldots,N\}$
with the uniform distribution and $V\dfn V_J$. Thus, our first necessary
condition for security level $h'$ is that there exists a random variable
$V$ with alphabet $\calV$ (jointly distributed with $U$) such that
$h'\le H(V)$. The second chain of inequalities is as follows:
\begin{eqnarray}
N(h'-\epsilon)&\le& H(V^N|W^m)\nonumber\\
&=& H(V^N|W^m,Y^n)+I(Y^n;V^N|W^m)\nonumber\\
&=& I(Y^n;V^N|W^m)\nonumber\\
&\le& H(Y^n|W^m)\nonumber\\
&\le& H(Y^n)\nonumber\\
&\le& \sum_{i=1}^nH(Y_i)\nonumber\\
&=& nH(Y|J')\nonumber\\
&\le& nH(Y),
\end{eqnarray}
where $J'$ is random variable taking values in the set $\{1,\ldots,n\}$
with the uniform distribution and $Y\dfn Y_{J'}$. The second equality is
due to the fact that $V^N$ is a function of $(W^m,Y^n)$ and so $H(V^N|W^m,Y^n)=0$.
Thus, another necessary
condition is the existence of random variable $Y$ at the output of
the channel $P_{Y|X}$ (which means the existence of a channel input variable
$X$ that induces $Y$ via $P_{Y|X}$) such that $h'\le \lambda H(Y)$. The combination of
the two necessary conditions then gives $h'\le\min\{H(V),\lambda H(Y)\}$.

A restatement of the necessity part of Theorem 1 would then be the following:
If $(D,R_c,h,h')$ is achievable then there exist a channel $P_{V|U}$
and a source $P_X$ such that the following conditions are simultaneously
satisfied:
\begin{itemize}
\item [(a)] $h \le H(U)-[I(U;V)-\lambda I(X;Y)]_+$,
\item [(b)] $h' \le \min\{H(V),\lambda H(Y)\}$,
\item [(c)] $R_c \ge I(U;V)$,
\item [(d)] $D \ge Ed(U,V)$.
\end{itemize}
Note that in contrast to Theorem 1,
we are no longer taking the minimum of $I(U;V)$ to obtain
$R(D)$, nor do we take the maximum of $I(X;Y)$ to obtain $C$. The reason
is that such optimizations might be in partial conflict with the
need to achieve large values of $H(V)$ and $H(Y)$ in order to meet condition
(b). Thus, there are more complicated compromises in the choice
of $X$ and $V$ when the tradeoff involves the additional parameter $h'$.

The achievability of this set of conditions remains open 
in general. However, for the special case where the channel $P_{Y|X}$
is deterministic, that is, $Y$ is a deterministic function of $X$,
and so $I(X;Y)=H(Y)$, the achievability scheme is 
essentially the same as before 
(but with general choices of $P_X$ and $P_{V|U}$) as long
as the required security $h'$ does not exceed the level
$\min\{I(U;V),\lambda I(X;Y)\}=
\min\{I(U;V),\lambda H(Y)\}$. If it is higher, and if $\lambda H(Y)$ exceeds $I(U;V)$
the additional key bits beyond $NI(U;V)$ (but not more than $H(V)$)
conveyed by the channel can be used to
control the (secret) choice of the rate--distortion codebook among
up to $2^{NH(V|U)}$ distinct codebooks 
that exist (cf.\ \cite{MM05},\cite{Merhav05})
and thereby achieve the extra security needed with regard to $V^N$.

Note that here, the separation principle no longer holds as before, in the
strong meaning of this term, because now, the choice of $P_X$ and $P_{V|U}$
involves compromises where there is an
interaction between the source coding of $U^N$ 
and the channel coding of $\bK$.

\subsection{Feedback}

Finally, consider the scenario of the previous subsection,
where in addition, there is noiseless feedback
from the channel output to the transmitter. 
In this case, it is clear too how to
secure $V^N$ to the level of 
$h'=\min\{H(V),\lambda H(Y)\}$, and it is also clear
that this value cannot be further improved upon. Here,
the encoder and the decoder simply share identical copies of $\{Y_i\}$
as a common key at both ends, and there is no longer use for the original
key, $\{K_i\}$. By the same token, in 
this case, the equivocation of $U^N$ can be enhanced to
the level of $h=H(U)-[I(U;V)-\lambda H(Y)]_+$, 
but not more. Thus, although feedback does not increase
the capacity of a DMC, it certainly improves 
its effectiveness when this channel serves
for key delivery.

\subsection{Continuous Alphabets}

In our derivations this far, we have limited ourselves to finite alphabet
sources and channels, primarily for reasons of convenience. Theorem 1
extends quite straightforwardly to the continuous alphabet case as well.
One comment is in order, however: In the continuous alphabet case, it no longer
makes sense to measure equivocation in terms of conditional (differential)
entropy, which can be negative.
It still makes sense, nonetheless, 
to define it by the complementary quantity - the
mutual information, $I(W^m;U^N)$, which is always non--negative.
Thus, part (a) of Theorem 1 would be
restated to assert that $[R(D)-\lambda C]_+$ is
an achievable lower bound to $I(W^m;U^N)/N$.

\newpage
\begin{figure}[h]
\hspace*{-2cm}\input{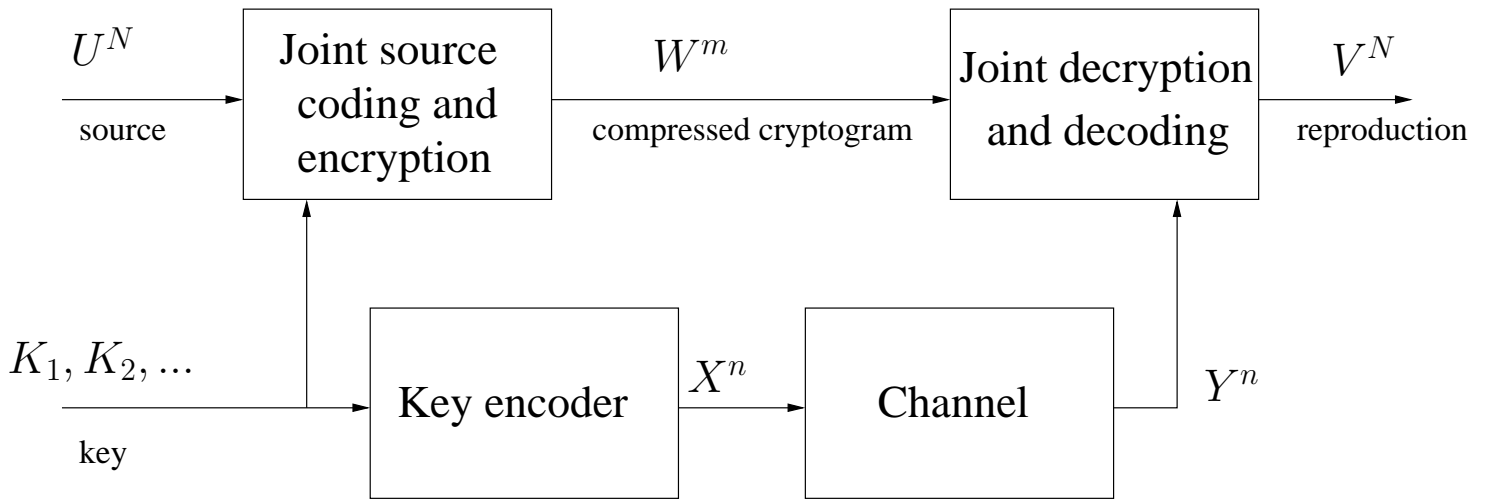}
\caption{A cipher system with capacity--limited key distribution.}
\label{gen}
\end{figure}
\end{document}